# Dynamic Wavelength-Tunable Photodetector Using Subwavelength Graphene Field-Effect Transistors

*François Léonard\*[†], Catalin D. Spataru[†], Michael Goldflam[‡], David W. Peters[‡], Thomas E. Beechem[‡]*

[†]Sandia National Laboratories, Livermore, CA, 94551, United States
[‡]Sandia National Laboratories, Albuquerque, NM, 87185, United States
*fleonar@sandia.gov

Dynamic wavelength tunability has long been the holy grail of photodetector technology. Because of its atomic thickness and unique properties, graphene opens up new paradigms to realize this concept, but so far this has been elusive experimentally. Here we employ detailed quantum transport modeling of photocurrent in graphene field-effect transistors (including realistic electromagnetic fields) to show that wavelength tunability is possible by dynamically changing the gate voltage. We reveal the phenomena that govern the behavior of this type of device and show significant departure from the simple expectations based on vertical transitions. We find strong focusing of the electromagnetic fields at the contact edges over the same length scale as the band-bending. Both of these spatially-varying potentials lead to an enhancement of non-vertical optical transitions, which dominate even the absence of phonon or impurity scattering. We also show that the vanishing density of states near the Dirac point leads to contact blocking and a gate-dependent modulation of the photocurrent.

1. **Introduction**

Photodetectors based on graphene have been extensively studied both experimentally[1-6] and theoretically[7,8]. A number of detector designs harnessing different photocurrent mechanisms have been realized and have shown promising performance. This success brings some interesting questions on whether graphene could provide novel modalities for photodetectors beyond the conventional photoresponse. One such modality is the ability to dynamically tune the spectral sensitivity of the detector. In existing technology, multi-color detection is achieved by using mechanical filters[9], multiple co-located pixels[9], or by using stacked layers[10], all of which have limitations: For example, mechanical filters are slow and introduce unwanted moving parts. In the multi-pixel approach, sub-pixels possessing differing spectral sensitivity are realized by changing material composition during fabrication. While providing hyperspectral response, resolution at a given wavelength is compromised. An alternative design is to stack multiple layers of materials possessing different absorption properties; while this approach is commonly used for two-color detection, generalizing the concept beyond two colors is difficult due to the necessity of monolithically integrating multiple materials and electronically addressing each layer. Furthermore, these methods do not allow for post-fabrication tuning of the spectral band. Thus, a photodetector concept with dynamic tuning of the intrinsic spectral range of a detector pixel composed of a single material possesses advantages over existing technology.

Graphene opens interesting possibilities to realize this concept, as recent work has shown that the *optical* properties of graphene can be modulated using a back gate[11,12] to realize optical modulators. If this result can be translated into modulation of the *photocurrent*, it could be harnessed for tunable photodetector applications. Unfortunately the experimental realization of this concept remains elusive, and to our knowledge the only experimental report of *spectral* photocurrent tunability (by a few tens of nanometers) was realized by inducing conformal changes in graphene[13]. In light of the experimental difficulties and non-idealities implicit in the fabrication of tunable graphene photodetectors, modeling can provide possible paths and identify key roadblocks. Previous modeling efforts have considered different possibilities for tunable graphene photodetectors based on *multiple* graphene layers, such as bilayer graphene photogating[14] and tunneling between two separate graphene layers[15] or nanoribbons[16]. These exciting results, based on macroscopic transport equations appropriate for long-channel devices, show the promise of devices based on multiple graphene layers. However, since devices based on monolayer graphene are easier to fabricate, an open question is whether wavelength tunability is possible in that case.

In this paper, we address the challenges of identifying promising tunable detector designs and the necessity for more detailed quantum modeling by developing and implementing a non-equilibrium quantum transport approach for photocurrent calculations in graphene devices, including realistic electromagnetic fields. Our main result is that dynamic hyperspectral imaging should be possible in such devices, but a wealth of phenomena renders the system rather complex. For example, we find that translational symmetry breaking due to spatially-varying band-bending and electromagnetic fields leads to a dominance of non-vertical transitions that significantly change the behavior expected from pure vertical transitions and Fermi blocking[17]. We also show that the vanishing density of states at the graphene Dirac point leads to a gate-dependent modulation of the photocurrent due to a contact blocking effect.

## 2. Methods

We consider a subwavelength graphene field-effect transistor (FET) as illustrated in Figs 1a,b. We focus on FETs with short channels for several reasons. First, the recombination time in graphene is on the order of picoseconds[18] which, combined with the Fermi velocity, gives a recombination length of about one micron. Therefore, at low light intensity we can focus on single electron-photon scattering events without recombination. Second, the subwavelength channel leads to a non-uniform vector potential that introduces novel optical transition phenomena. Finally, because we consider uniform illumination of the full device, no temperature gradient exists between the two contacts and photothermoelectric effects[19] can be neglected. While we present results for hot electron devices (i.e. purely ballistic transport), our results can be directly converted to the scattering case by multiplying the calculated photoresponse by the ratio $l_{mfp}/L$ where $l_{mfp}$ is the electron scattering mean free path and $L$ is the channel length.

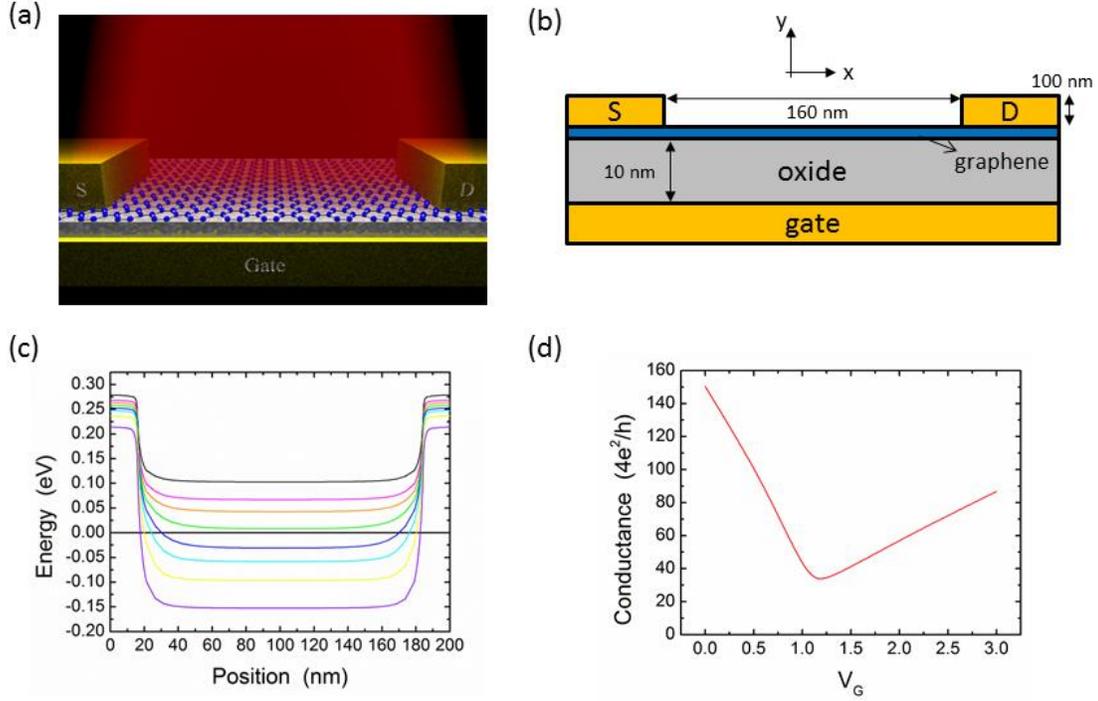

**Figure 1.** (a) Schematic of the photodetector under consideration. A single sheet of graphene is contacted by source and drain electrodes, and is separated from the gate electrode by a dielectric. Monochromatic light illuminates the device. (b) Cross-section of the graphene device with the coordinate system and the dimensions used in the simulations. The oxide is lossless and dispersionless with a dielectric constant of 3.9. (c) Position of the graphene Dirac point along the length of the FET. Curves from top to bottom correspond to gate voltages of 0, 0.5, 0.75, 1, 1.25, 1.5, 2, and 3 V. (d) Dark transfer characteristic at room temperature computed from the band-bending of panel (c).

To calculate the photocurrent, we employ the non-equilibrium Green's function (NEGF) approach, where we explicitly include the electron-photon interaction at the quantum level. We previously developed and applied this approach to carbon nanotube p-n junctions[20]; the graphene FET case differs significantly not only because of the material and the device geometry, but also because the FET does not possess a built-in field to generate a photocurrent at zero bias. Furthermore, to our knowledge, the only NEGF photocurrent calculation for graphene presented in the literature[21] (for a model p-n junction) did not include a non-uniform electromagnetic potential as considered in this paper.

Our calculations consist of two preliminary steps and a final step to obtain the photocurrent. The first preliminary step is a self-consistent NEGF calculation of the potential and charge in the *dark* graphene FET using our previously developed approach for carbon nanotubes[20,22,23] appropriately modified for graphene (specific details can be found in the Supporting Information). In the second preliminary step, we solve Maxwell's equations in the geometry of Figs 1a,b to obtain the vector potential on the graphene sheet. The final step is to combine the vector potential and the NEGFs from the dark simulations to calculate the photoresponse. Because the graphene FET does not possess a net built-in field, no photocurrent is generated for uniform illumination at zero bias. However, a small source-drain bias breaks the symmetry and leads to a net photocurrent; we therefore focus on the photoconductance in the small bias limit, defined as

$$\sigma^{ph} = \frac{\partial I_{ph}}{\partial V_{sd}}\bigg|_{V_{sd}=0} \tag{1}$$

where $I_{ph}$ is the photocurrent at a source-drain bias $V_{sd}$. The full expression for $\sigma^{ph}$ can be found in the Supporting Information.

In our simulations, we model the graphene using a tight-binding Hamiltonian with overlap integral of 2.5 eV and nearest-neighbor bond length of 0.142 nm. The zigzag direction of the graphene is perpendicular to the electron transport direction. The source and drain electrodes are perfect metals vertically separated from the graphene sheet by a distance of 0.3 nm. Charge transfer between the metal and the graphene is naturally captured in our simulations by using a metal workfunction 0.5 eV larger than graphene. The electron-photon interaction is included by modifying the tight-binding Hamiltonian with the usual $\vec{A}\cdot\vec{p}$ term where $\vec{A}$ is the magnetic vector potential of the light field in the device geometry at the graphene location and $\vec{p}$ is the graphene electron momentum.

## 3. Results

Figure 1c shows the calculated self-consistent band-bending potential for different gate voltages ranging from 0V to 3V (the band-bending potential is equivalent to the energy of the Dirac point). As expected for a FET, the potential in the middle of the channel is modulated by the gate; however, the metal contacts tend to pin the potential, leading to a band-bending over a distance of ~ 10 nm near the contacts. This length scale is determined by the screening environment of the device and the graphene. In this case, the bottom metal gate provides most of the screening, and the band-bending length scale is determined by the thickness and dielectric properties of the oxide. An additional effect is seen where the band potential under the contacts also changes with the gate bias. This gate modulation of the contacts has been previously discussed in the case of carbon nanotubes[24].

Figure 1d shows the small bias conductance in the dark based on the band-bending potential of Fig. 1c which displays an asymmetric shape and a minimum conductance of about 30 times the quantum of conductance of $4e^2/h$. This result is in good agreement with experimental data for short channel devices[25,26] which come within a factor of three of the minimum conductivity calculated here, and also show prominent asymmetry. In our case, the asymmetry is due to the contact modulation by the gate.

We obtained the vector potential $\vec{A}$ by solving Maxwell's equations in the device geometry of Fig. 1a assuming a periodic structure with perfect metals for the contacts and the bottom gate, for a monochromatic plane wave normally incident and polarized in the x direction (see Supporting Information for details of the calculations). Figure 2a shows that interaction of the plane wave with the device leads to the appearance of an additional component of the vector potential in the y direction. More importantly, the vector potential exhibits strong spatial dependence with significant enhancement near the contacts. Figure 2b shows the ratio of $A_x$ at the contacts to that in the middle of the device, which is about 36 at the longer wavelengths and increases to more than 60 at a wavelength of 1 micron. These results suggest that the properties of photodetectors based on two-dimensional materials should be particularly sensitive to near-contact effects even under flood illumination.

The near-contact focusing of the electromagnetic field is spatially co-located with the device band-bending. As discussed in the context of Fig. 1c, the band-bending near the contact occurs over a distance ~ 10 nm and is determined by the dielectric thickness. We find that the electromagnetic field also decays over this same length scale as illustrated in Fig. 2c. We believe this is also a consequence of the screening in the device geometry. For example, when solving Maxwell's equations, the boundary conditions at the source and drain electrodes can be satisfied by introducing surface charges and currents, and these will in turn be screened by the bottom gate. We verified this idea by repeating our simulations for different oxide thicknesses, finding that the vector potential decay length (defined as the 1/5 decay distance from the contact) increases with increasing oxide thickness (Fig. 2d).

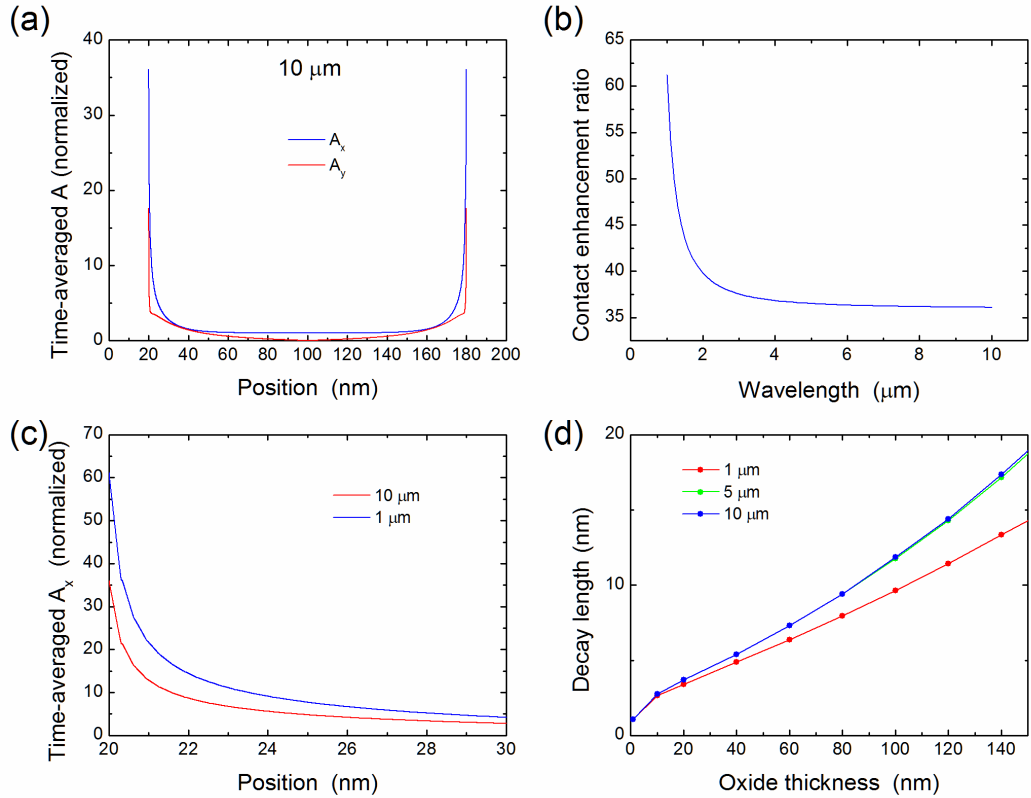

**Figure 2.** (a) Components of the time-averaged vector potential at the graphene sheet for an incoming monochromatic light wave of 10 µm wavelength and polarized in the x direction. The vector potential is normalized to a value of 1 in the middle of the channel. (b) Ratio of $A_x$ near the contact to that in the middle of the channel as a function of wavelength. (c) Vector potential near the left contact for two different wavelengths of the incoming light. Incoming light polarization is along the x axis. (d) Dependence of the decay length of the vector potential as a function of the oxide thickness.

Having obtained the gate-dependent dark band-bending and the vector potential resulting from illumination, we can calculate the photoconductance. Since the optical absorption in graphene for the out-of-plane contribution is very small[27], we can neglect the $A_y$ component and consider only the spatially-varying $A_x$ component. Figure 3 shows the results of such calculations for two photon energies of 0.1 eV (12.4 µm) and 0.25 eV (4.95 µm). Several conclusions emerge from the data in this figure. First, the photocurrent is gate-tunable for each photon energy. Second, there are regions where clear discrimination between wavelengths is possible. Third, the curves do not follow the simple expectation based on Fermi blocking of vertical transitions. Indeed, if only vertical interband transitions are allowed, the symmetric graphene bands imply that the Fermi level must be less than $\hbar\omega/2$ above or below the Dirac point for transitions to occur. Using the Dirac point energy in the middle of the graphene FET to extract the gate voltage at which transitions should be blocked, we obtain the vertical dashed lines in Fig. 3, outside of which no photocurrent should be observed. The full calculations show a significant deviation from this simple model for both photon energies.

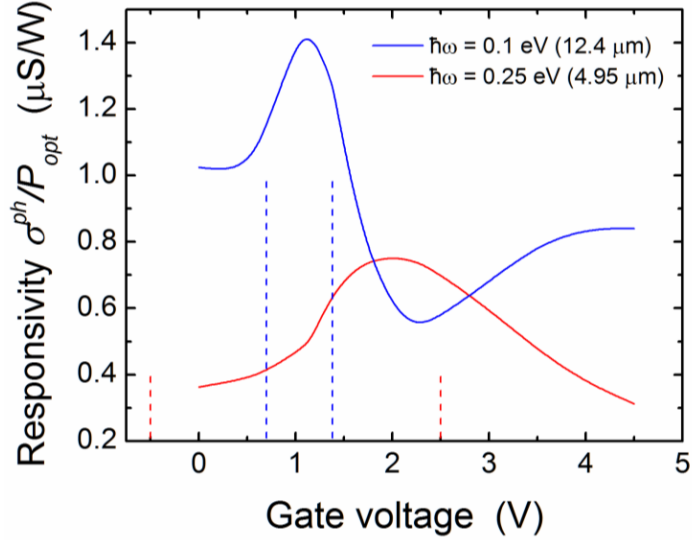

**Figure 3.** Photoconductance responsivity $\sigma^{ph}/P_{opt}$ as a function of gate voltage, calculated for two different photon energies. The vertical dashed lines are the boundaries outside of which Fermi blocking would be expected to cut-off the photoresponse.

The above effects have their origin in two main factors: non-vertical transitions and gate-dependent contact blocking. Non-vertical transitions can arise when the translational symmetry is broken; in the case of the graphene FET translational symmetry is broken by the spatially-varying electrostatic potential and spatially-varying magnetic vector potential. The missing momentum that allows non-vertical transitions arises from the infinite number of Fourier modes that make up these spatial variations. To illustrate the importance of non-vertical transitions, we plot in Fig. 4 the effective transmission probability for electrons as a function of their energy compared to the band-bending in the graphene FET at a gate voltage of 4.5 V. This transmission represents the probability that an electron injected in the source at energy $E - \hbar\omega$ absorbs a photon and is collected in the drain at energy $E$. If only vertical transitions were present, the transmission would show a peak at energy $E = \hbar\omega/2$ above the Dirac point. (Note that because of the occupancy of the states, at zero temperature the transmission would be non-zero only in an energy window between $E_F$ and $E_F + \hbar\omega$. Thermal broadening extends this window by a few $k_B T$ above and below.)

For the case of Fig. 4, the Dirac point in the middle of the channel is more than 0.2 eV below the Fermi level. In principle the photoresponse should vanish unless the photon energy is larger than about 0.4 eV. However, we find that for photon energies of 0.1 eV and 0.25 eV, a significant photoresponse remains (Fig. 3). This can be explained from examination of Fig. 4b,d where the interband vertical transitions are Fermi-blocked in both cases but significant transmission is observed at energies where only non-vertical transitions can arise. For example, at $\hbar\omega = 0.1$ eV vertical interband transitions would give a peak at -0.17 eV based on the mid-channel Dirac point, but these transitions are blocked because the Fermi level is at 0 eV and there are no empty states to excite into. However, non-vertical transitions can exist between $\approx$-0.1 eV and $\approx 0.2$ eV (i.e. the Fermi window plus thermal broadening, and with appropriate momentum transfer from spatial symmetry breaking) in agreement with the transmission. By increasing the photon energy to $\hbar\omega = 0.25$ eV we see the appearance of another effect: contact blocking. Indeed, at this photon energy, non-vertical transitions exist up to energies of 0.25 eV (plus thermal broadening). However, the transmission function has a strong dip to zero in the middle of this range at an energy that corresponds to the Dirac point in the contacts. This strong reduction in transmission is a consequence of the density of states of graphene that goes to zero at the Dirac point; thus, one can think of this effect as contact blocking of the photocurrent.

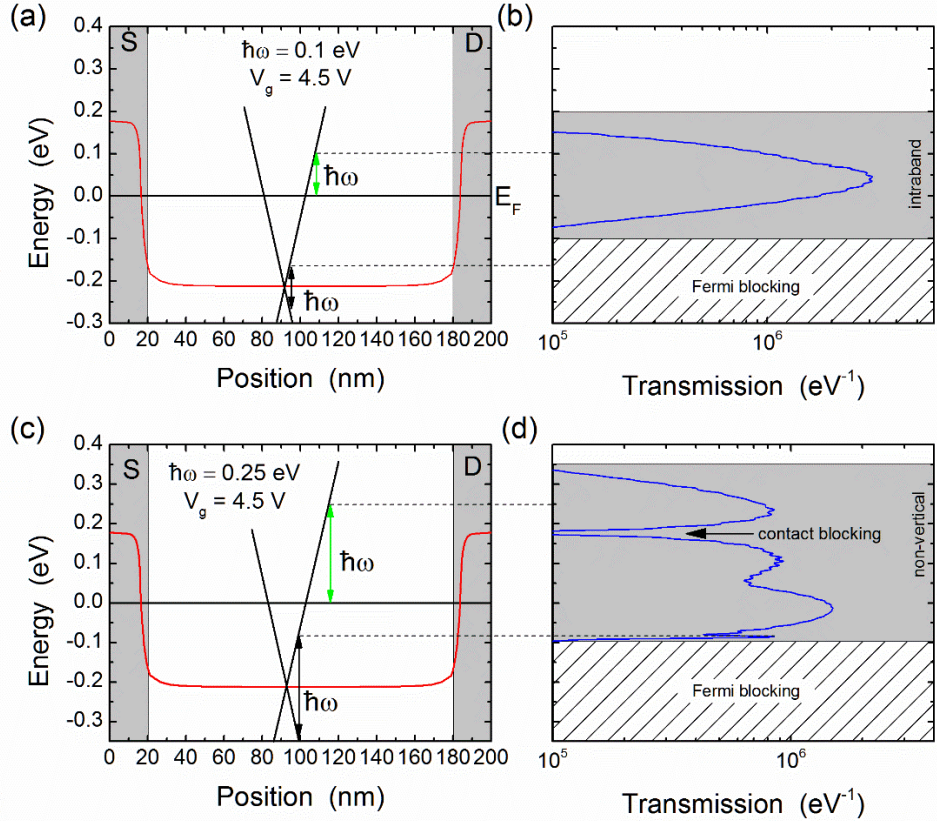

**Figure 4.** Comparison between the band-bending and the effective transmission probability for photo-excited carriers. The top panels (a,b) are for a photon energy of 0.1 eV while the bottom panels (c,d) are for a photon energy of 0.25 eV. In both cases the gate voltage is 4.5 V. The vertical grey-shaded regions in the left panels are the source (S) and drain (D) contacts. For illustration purposes, we have also sketched the graphene crossing bands with Dirac point in the middle of the channel. The black vertical arrowed lines denote interband transitions of energy $\hbar\omega$.

This effect is in part responsible for the shape of the photoresponse versus gate voltage observed in Fig. 3. To illustrate this effect in more detail, Fig. 5 plots the band-bending and transmission probabilities for a gate voltage of 1 V. In the case of $\hbar\omega = 0.1$ eV, a small peak at 0.061 eV is visible due to vertical transitions, but most of the photoresponse comes from non-vertical transitions. A similar situation arises for $\hbar\omega = 0.25$ eV; however, the Dirac point in the contact regions is now located in the Fermi window and blocks a significant portion of the carriers, leading to a reduced photoresponse and the unusual shape observed in Fig. 3.

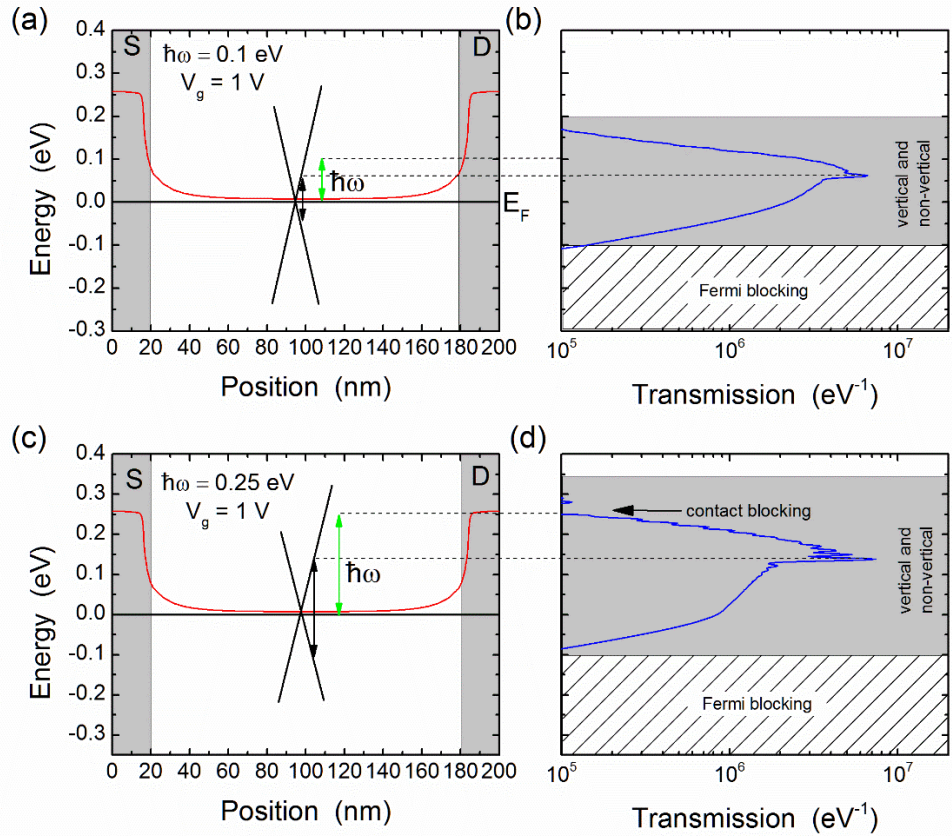

**Figure 5.** Comparison between the band-bending and the effective transmission probability for photo-excited carriers. The top panels (a,b) are for a photon energy of 0.1 eV while the bottom panels (c,d) are for a photon energy of 0.25 eV. In both cases the gate voltage is 1 V. The vertical grey-shaded regions in the left panels are the source (S) and drain (D) contacts. For illustration purposes, we have also sketched the graphene crossing bands with Dirac point in the middle of the channel. The black vertical arrowed lines denote vertical interband transitions of energy $\hbar\omega$.

To further determine the factors that lead to the unusual importance of non-vertical transitions, we calculated the transmission for a gate voltage of 1 V and illumination at $\hbar\omega = 0.1$ eV. For this gate voltage, the Dirac point is 0.011 eV above the Fermi level in the middle of the channel and vertical interband transitions should give a peak at an electron energy of 0.061 eV. As shown in Fig. 6, a peak is observed at this energy, but is significantly broadened by non-vertical transitions. We then repeated this calculation for a constant electrostatic potential throughout the device (i.e. flat bands) with a value corresponding to the Dirac point in the middle of the channel for the full device simulations at $V_g = 1$ V. Figure 6 shows that at low temperatures (30K in our case) and for a uniform vector potential in the channel, there is a narrow peak due to vertical interband transitions; increasing the temperature to 300K does not change this behavior and we therefore rule out temperature effects as the main source of the non-vertical transitions. However, introducing a non-uniform vector potential (but still with flat bands) gives a much broader peak, implying that the non-uniform light profile itself leads to significant non-vertical transitions. Finally, since the full device band-bending gives an even broader peak, non-vertical transitions are also enhanced by the spatially-varying band-bending that breaks the spatial symmetry and provides the momentum.

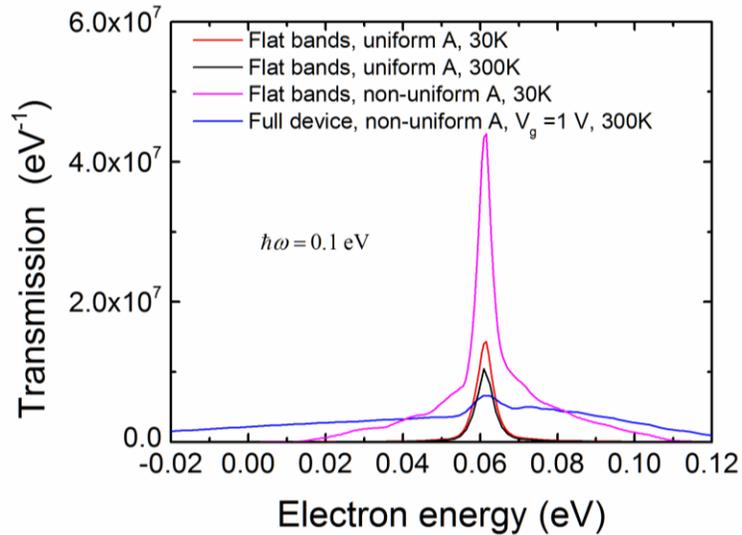

**Figure 6.** Electron transmission for a photon energy of 0.1 eV and a full device band-bending for a gate voltage of 1V. The result is compared with different flat band cases.

## 4. Conclusion

In conclusion, we find that monolayer graphene FETs open a new avenue to realize photodetectors with dynamic spectral tunability, and display a number of novel underlying phenomena that should be general. For example, electromagnetic field focusing at the contacts plays a key role in governing photoresponse behavior, a phenomenon that should apply to a broad range of one- and two-dimensional materials. We also expect that non-vertical transitions induced by symmetry breaking due to the spatial variations of the band-bending and the vector potential should also become prominent in other low-dimensionality materials. Finally, additional studies are needed to assess the robustness of the wavelength tunability to non-ideal factors such as substrate charged impurities. Nonetheless, our work establishes the first framework to guide and interpret experimental and theoretical efforts.

**Acknowledgements**

Work supported by the Laboratory Directed Research and Development program at Sandia National Laboratories. Sandia National Laboratories is a multi-program laboratory managed and operated by Sandia Corporation, a wholly owned subsidiary of Lockheed Martin Corporation, for the U.S. Department of Energy's National Nuclear Security Administration under contract DE-AC04-94AL85000. We thank Dr. Aron Cummings for help with the computational NEGF code.

**Author contributions**

All authors contributed to development of the project and writing of the manuscript. F.L., C.S., and M.G. performed the computational work.

**Competing financial interests**

The authors declare no competing financial interests.

# Dynamic Wavelength-Tunable Photodetector Using Subwavelength Graphene Field-Effect Transistors

*François Léonard*[*,†], *Catalin D. Spataru*[†], *Michael Goldflam*[‡], *David W. Peters*[‡], *Thomas E. Beechem*[‡]

[†]Sandia National Laboratories, Livermore, CA, 94551, United States

[‡]Sandia National Laboratories, Albuquerque, NM, 87185, United States

*fleonar@sandia.gov

**Supporting Information**

*1. Overall theoretical/computational approach*

Figure S1 shows the overall approach for calculating the photocurrent. The first two isolated steps consist in the calculation of the dark band-bending and the light fields in the device geometry. The third step is a one-shot calculation of the photocurrent using input from the first two steps.

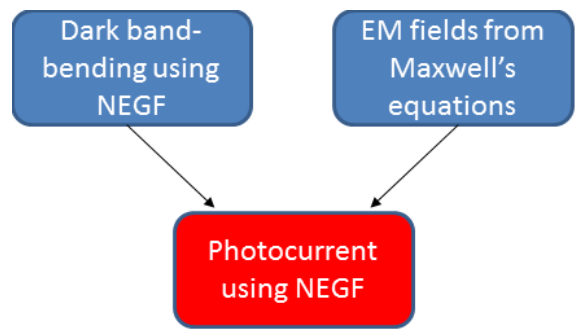

**Figure S1.** Illustration of the general approach to calculate the photocurrent.

*2. Calculation of dark band-bending*

Our approach to calculate the band-bending has been documented extensively in the case of carbon nanotubes[22]. We describe it here paying attention to the details specific to graphene. We calculate the dark band-bending by performing a self-consistent calculation between the electrostatic potential and the graphene charge. The electrostatic potential is obtained by solving Poisson's equation $\nabla \cdot (\varepsilon(x,y,z) V(x,y,z)) = -\rho(x,y,z)$ with the source charge $\rho(x,y,z)$ being the graphene charge. The coordinate system is given in Figure 1 of the main text. The dielectric constant varies spatially in the y direction at the interface between the oxide and the vacuum above it. Fixed values of the potential are set at the source, gate, and drain electrodes. At the left, right, and top of the simulation cell, we set a boundary condition of zero derivative on the electrostatic potential. The vacuum space above the device is chosen large enough that the results are converged with respect to that height. Computationally, we use an in-house custom code based on a non-uniform grid to solve Poisson's equation. The calculation is accelerated by using a periodic unit cell in the z direction, and using a fast Fourier transform algorithm for that direction.

At the boundaries between the graphene and the contacts or the oxide, the graphene is vertically separated by a van der Waals gap of 0.3 nm on either side. We have discussed the modeling of contacts in this manner previously[24].

Once the electrostatic potential is obtained, we use the non-equilibrium Green's function approach to obtain the charge on the graphene. (Since we only need the zero bias band-bending, in practice the calculations are accelerated by only using equilibrium Green's function.). This is done within a tight-binding model with nearest-neighbor interaction of value 2.5 eV. The graphene atoms are located at vertical position $y = h$, and the electrostatic potential is evaluated at the location of the graphene atoms $V(x_i, h, z_i)$ where $(x_i, h, z_i)$ is the location of atom *i*. We then add $-eV(x_i, h, z_i)$ as a diagonal term in the tight-binding Hamiltonian. The total charge on each graphene atom is obtained by solving the usual NEGF equations with semi-infinite graphene leads. In practice, the calculations are performed by considering a graphene nanoribbon of index (n,m) corresponding to the notation familiar from carbon nanotubes. In our case, the zigzag edge is parallel to the contact edges, and therefore the nanoribbon is of (n,0) type. We solve the NEGF equations independently for each band $p = 1,...,n$ and sum the total charge from each band. We found that $n = 100$ gives good convergence of the total charge.

Once we have the total charge on each atom, we use a 3D Gaussian profile at each atom site to distribute the charge and create the charge density $\rho(x, y, z)$ that enters Poisson's equation. The electrostatic potential is re-calculated, and used as the new input to the charge calculation. This process is repeated until self-consistency is achieved using the Broyden algorithm.

*3. Calculation of electromagnetic fields*

Using COMSOL Multiphysics, we obtain the electromagnetic fields by solving Maxwell's equations for a periodic array, with the device geometry shown in Fig. 1a of the main text. The device is excited by a uniform plane wave polarized in the x direction. The source, drain, and gate contacts are modeled as perfect electric conductors and the fields at the device edges are forced to satisfy periodic boundary conditions. For the results in the main text, a non-dispersive model of a generic oxide is employed with $\epsilon_1=3.9$ and $\epsilon_2=0$. Using this simplified optical model yields more generally applicable results, which may be extended to a range of more complex dielectric materials.

To examine the effects of using more realistic oxide properties, we ran identical simulations employing a fully dispersive $SiO_2$ model. The optical properties employed, as well as the resulting contact enhancement ratio, are shown in Figure S2. The enhancement ratio shows a strong dependence on the permittivity of $SiO_2$, suggesting the important role of the dielectric in the device optical response. However, the enhanced electric fields at the contacts is still a general feature across the wavelength range.

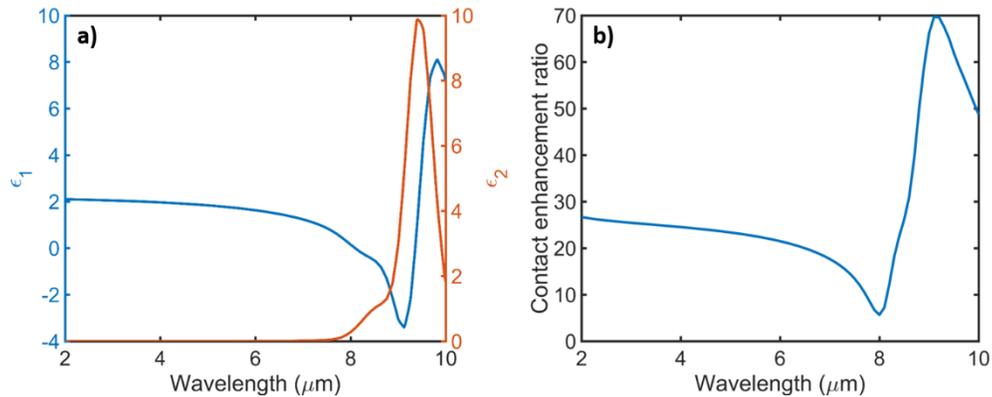

**Figure S2.** a) Real ($\epsilon_1$) and imaginary ($\epsilon_2$) $SiO_2$ permittivity employed in COMSOL modeling. b) Contact enhancement ratio obtained from simulations using realistic $SiO_2$ optical properties.

4. Calculation of photocurrent

We calculate the photocurrent using the NEGF approach we previously developed in the context of carbon nanotubes[20], but modified for the particular case of the graphene FET. For small light intensities, and in the absence of multi-photon processes, current in a device can be obtained from the expression

$$I = \frac{4e\gamma}{h}\int dE \, \text{Re}\left\{G^{<(dark)}_{0,1}(E) + G^{<(dark)}_{N,N+1}(E)\right\} + \frac{4e\gamma}{h}\int dE \, \text{Re}\left\{G^{<(ph)}_{0,1}(E) + G^{<(ph)}_{N,N+1}(E)\right\} \quad (2)$$

where $G^{<(dark)}_{i,j}$ is the Hamiltonian Green's function between tight-binding sites i and j in the absence of light, and $G^{<(ph)}_{i,j}$ comes from the linear expansion in the photon flux. The Green's functions in the dark correspond to those used when obtaining the self-consistent charge and potential in the device geometry, as discussed in section 2.

The photocurrent is defined as

$$I^{(ph)} = \frac{4e\gamma}{h}\int dE \, \text{Re}\left\{G^{<(ph)}_{0,1}(E) + G^{<(ph)}_{N,N+1}(E)\right\} \quad (3)$$

with

$$G^{<(ph)} = G^{R0}\Sigma^{<(ph)}G^{R0\dagger} + G^{R0}\Sigma^{R(ph)}G^{R0}\Sigma^{<0}G^{R0\dagger} + G^{R0}\Sigma^{<0}\left(G^{R0}\Sigma^{R(ph)}G^{R0}\right)^{\dagger}. \quad (4)$$

In this equation $G^{R0} = \left[EI - H_0 - \Sigma^{R0}\right]^{-1}$ with $H_0$ the bare Hamiltonian and $\Sigma^{R0}$ the self-energy due to the semi-infinite graphene leads. For the dark state we also have $\Sigma^{<0} = -2f \, \text{Im}\Sigma^{R0}$ with $f$ the Fermi distribution function.

The electron-photon interaction is captured through the function

$$\Sigma^{\{R,<\}(ph)}(E) = \alpha \sum_{pq} P_{lp}P_{qm}G^{\{R,<\}0}_{pq}(E - \hbar\omega) \quad (5)$$

with

$$P_{lm} = \delta_{l\pm 1,m}\left[f^{\pm}(l) + f^{\pm}(m)\cos\left(\frac{\pi J}{M}\right)\right]. \quad (6)$$

In these equations, $f^{\pm}(l) = \mp 1 - (-1)^l$ and $\alpha = \frac{e^2 a^2 \gamma^2 F}{2\hbar\omega c\varepsilon}$ with $a = 0.07$ nm the smallest separation between adjacent lines of carbon atoms perpendicular to the transport direction, $\gamma = 2.5$ eV the tight-binding overlap integral, F the photon flux, $\hbar\omega$ the photon energy, c the speed of light, and $\varepsilon$ the permittivity of free space.

Equation (5) is obtained by considering a graphene channel of finite width but with periodic boundary conditions in the direction perpendicular to the transport direction. In this case, there are *M* bands each described by a quantum number *J*. In practice, we have found that the photocurrent is converged by using ~500 bands.

In the case of the graphene FET device, the absence of an asymmetric built-in field implies that the photocurrent is zero unless a finite bias is applied. It is certainly possible to perform the NEGF calculations in the presence of a finite bias, but since the response is linear at small bias we can take advantage of the fast numerical calculations of the dark Green's functions and focus on the *photoconductance*, defined as

$$\sigma^{ph} = \left.\frac{\partial I^{(ph)}}{\partial V_{sd}}\right|_{V_{sd}=0} \tag{7}$$

We note that previous theoretical work[28] has demonstrated that such linear response functions in the presence of arbitrary interactions can be written as

$$\sigma^{ph} = \frac{4e^2\gamma}{hk_BT}\int dE\, T_{eff}(E) \tag{8}$$

where $T_{eff}(E)$ is an effective transmission function. In our case the effective transmission function can be written as

$$T_{eff}(E) = \left[\frac{f(E-\hbar\omega)-f(E)}{f(E-\hbar\omega)}\right]\Gamma(E)\text{Im}\left[G^{R0}(E)\Sigma^{<(ph)}G^{R0\dagger}(E)\right]_{NN}. \tag{9}$$

This is the transmission function that we plot in the main text.

*4. Responsivity*

Traditionally the responsivity is defined as

$$\text{responsivity} = \frac{\text{photocurrent}}{\text{optical power}} \tag{10}$$

but since we calculate the photoconductance instead of the photocurrent, we define the photoconductance responsivity as

$$\mathcal{R} = \frac{\sigma^{ph}}{P_{opt}} \tag{11}$$

where $P_{opt}$ is the total optical power incident on the graphene channel.